\documentclass[pageno]{jpaper}
\usepackage[pdf]{graphviz}
\usepackage{pgf-umlsd}
\usepackage{listings}
\usepackage{enumitem}
\usepackage{amsfonts}
\graphicspath{ {./figs/} }
\usepackage{framed,color,verbatim}
\usepackage{mathtools}
\definecolor{shadecolor}{rgb}{.9, .9, .9}

\newenvironment{code}%
   {\snugshade\verbatim}%
   {\endverbatim\endsnugshade}

\usepackage[normalem]{ulem}

\begin{document}

\title{A multi-chain synchronization protocol that leverage zero knowledge proof to minimize communication trust base.}
\author{Sinka Gao, Li Guo Qiang, Fu Hong Fei, Zhang Heng}
\newcommand{\dprotocol}{Delphinus Cross-Chain Aggregator }
\date{}
\maketitle
\thispagestyle{empty}
\begin{abstract}
\dprotocol is a universal firmware which synchronise states between different smart contracts on different block-chains. In the world of block-chains, synchronization challenges are two-folded. Firstly, contracts from different main block-chain can not communicate with each other which makes it hard to establish a trustworthy communication channel for them to share and maintain a universal state between each other. Secondly, transactions on different block-chains can hardly be ordered thus conflicts are common and we need a novel way to avoid and handle these conflicts. \dprotocol is a ZKSNARK based multi-block-chain layer on top of which rich cross chain applications can run safely and efficiently.
\end{abstract}

\section{Introduction}
As an emerging distributed computing paradigm, block-chain is rapidly evolving in areas such as digital finance and cryptography. However, existing block-chain projects adopt different block-chain architectures and protocols and, as a consequence, it is difficult in general for different block-chain systems to flow information to each other. Thus different block-chains themselves are born isolated islands which brought limitations to the overall usability, functionality and scalability of block-chain technology. \cite{anati2013innovative}

Recently, the growing demand of flowing value between different chains stimulates the demand of secure and consistent cross chain exchange protocols for cross-chain communication and book-keeping.  In order to address the safety, liveness, permissionless and linearizability problem of various protocols, and build trustworthy consensus between different block-chain networks, cross-chain techniques received a lot of attention.

In general the difficult part of cross-chain communication is that there is no proper way for a block-chain transaction to confirm whether a transaction was executed successfully in another block-chain. This is because block-chain system is a decentralized computing system that contains a large set of nodes connected over a peer-to-peer network and a successful transaction needs to be audited on multiple nodes, which means that a transaction's result can only depends on block-chain's internal states otherwise the result of a transaction will be inconsistent between different auditing nodes. 

This builtin drawback of block-chain, makes it hard to execute a group of transactions involving different chains so that either all of them or none of them succeed on different chains. For example, suppose that Alice would like to buy one token from Bob on block-chain A by spending one token on block-chain B. To achieve this swap, two transactions $tx_A$ and $tx_B$ are bundled:
\begin{itemize}[leftmargin=*]
\item $tx_A$: Alice get a token from Bob on block-chain $C_A$.
\item $tx_B$: Alice pay a token to Bob on block-chain $C_B$.
\end{itemize}
Moreover $tx_A$ and $tx_B$ need to be executed in a safe manner that either both of them succeed or both of fail. If it is not the case, then either Alice lose a token on block-chain B or Bob lose a token on block-chain A. However since $tx_A$ can not access any info of $tx_B$, certain protocol needs to be designed to achieve a safe swap.

Many protocols are developed in the literature including time hash lock, interledger’s stream protocol, relay system, spv, side chain, etc and most of them introduce a relayer to establish an information channel from source block chain A to target block chain B as follows:
\begin{figure}[h]
\caption{forward message using a relayer}
\label{relayer-connection}
\includegraphics[scale=0.6]{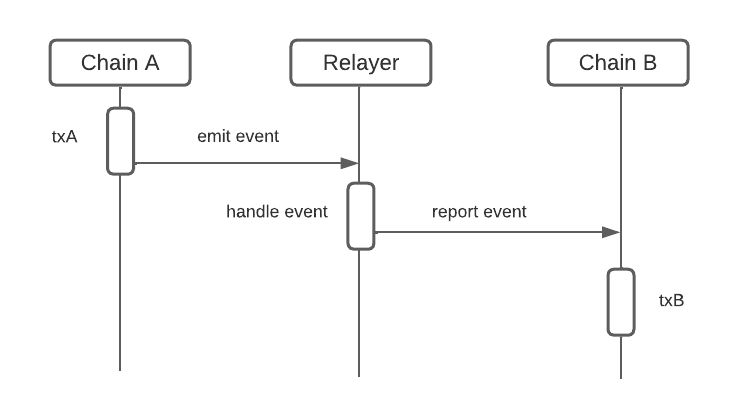}
\end{figure}

The major security problems regarding the above model is that we require the relayer to be honest and can be identified. A centralized relayer can be build such that chain B only accepts info reported from a trusted relayer. In such solution, the trusted relayer sign all their messages with their private key and $tx_B$ in chain B checks signature accordingly. The problem of a centralized relayer is that the relayer becomes one of the most important trust base for the protocol to work which may leads to SIOF(single point of failure) if relayer is hacked or misbehaved.

To take a step further, if we consider a protocol between Alice and Bob that involves three transactions $tx_A^0$ on block-chain $C_A$, $tx_B^1$ on block-chain $C_B$ and $tx_A^2$ on block-chain $C_A$ in the following order:
\begin{lstlisting}[escapeinside={(*}{*)}, backgroundcolor = \color{lightgray},]
  (*$s_A' = tx_A^0(s_A)$*);
  (*$s_B' = tx_B^1(x,s_B)$*);
  (*$s_A'' = tx_A^2(y, s_A')$*);
\end{lstlisting}
Then it follows that the safety property require transactions involved in the above bundle to be either all succeed or none of them succeed. Suppose a protocal is designed to help carrying out such bundled transactions between block-chains then we can define the safety, permissionless, liveness and linearizability properties of the protocol as follows:

\begin{enumerate}[leftmargin=*]
\item permissionless: Every nodes of certain block-chain are allowed to join the communication network.
\item safety: All state updates for a single transaction are either successful or fail on all involved block-chains.
\item liveness: All valid transaction will eventually success on all block-chains
\item linearizability: All parallel successful transactions is always linearlizable so that the consequent state of a group of parallel transactions $tx_i$ is equivalent to performing the transactions sequentially in a special order that respects the happen-before relation of the original order.
\end{enumerate}

\subsection{Preliminary}
\label{prelimiary}
While there are many existing protocols designed for specific scenarios and they focus on preserving different properties with various trust base. \dprotocol targets to achieve liveness, safety, permissionless and linearizability properties by using zero knowledge proof to provide a universal layer to synchronise states between different smart contracts on different block-chains and this layer together with its communication protocol can be used as a cross-chain communication framework. Below we will first clarify a few key definitions in our protocol and then describe the main idea behind \dprotocol.
\\
\subsubsection{Block-chain}
A block-chain is a distributed database that is shared among network of distributed nodes\cite{chen2018survey}. As a database, transactions are first submitted and organized in blocks, while each block is a data structure that records the most recent transactions which are not yet validated. Once a block was validated the state of the whole distributed database is updated accordingly.\\
\newline
We use $C$ to denote a particular block-chain system with distributed nodes storing data of $C$ in a distributed way. As the state of block-chain $C$ is updated base on the validation of blocks, the observable state of a block-chain can be abstracted as a state sequence. We use $S_k$ to denote the whole state of a block-chain $C_{k}$ and $s_k$ to denote a partial state of $S_k$.
\\
\subsubsection{Bundled cross-chain transaction}
Given a group of block-chain $\left\{C_{1}, C_{2}, \cdots, C_{k}\right\}$, we say a transaction $tx$ is a bundled cross-chain transaction if it is composed of a sequence of transactions $tx_{k_0}^0$, $tx_{k_1}^1$, $\cdots$ $tx_{k_j}^j$ such that $j$th involved sub-transaction $tx_{k_j}^j$ is supposed to be executed on $C_{k_j}$ locally and the behavior of each $tx_{k_j}^{j}$ depends on state $s_{k_j}$. \\
\newline
In a bundled transaction $tx$, each local transaction $tx_k^j$ is executed on the partial state of $s_k$. We use $s= \langle s_1, s_2, \cdots, s_k \rangle$ to denote the underlying global state of $tx$ and we use $s_k$ to denote the projection of $s$ on block-chain $C_k$. For example, a bundled transaction could looks like the following. 
\begin{lstlisting}[escapeinside={(*}{*)}, backgroundcolor = \color{lightgray}, basicstyle=\small,]
tx(s) {
    (*$s^1 = \langle tx_0(s_0^0), s_1^0, s_2^0, \cdots \rangle$*)
    (*$s^2 = \langle s_0^1, tx_1(s_1^1), s_2^1, \cdots \rangle$*)
    ....
    (*$s^n = \langle s_0^{n-1}, s_1^{n-1}, s_2^{n-1}, tx_n(s_{k}^{n-1})\rangle$*)
}
\end{lstlisting}
where $tx_0$ is performed on local state $s_0$, $tx_1$ is performed on local state $s_1$ ..., and the global state $s$ is changed to $s^n$ after $tx$ is fully performed.
\\
\subsubsection{Aggregator Chain}
In this work, we will introduce an extra block-chain to store the global state $s$. We denote this chain as an aggregator chain to distinguish between it from other block-chains $C_{k}$ involved in a bundled cross-chain transaction $tx$. Also we will use native-chain as an abbreviation of $C_{k}$ to distinguish them from our aggregator chain.
\\
\subsubsection{Merkely Tree Encoding of Global State.}
Merkle\cite{becker2008merkle} tree is a tree in which every leaf node is labelled with the cryptographic hash of a data block, and every non-leaf node is labelled with the cryptographic hash of the labels of its child nodes. Each data block has a unique Merkle tree index(MTI) and each MTI encodes a unique path from top to leaf.\\

\begin{figure}[!ht]
\caption{merkle tree structure}
\label{merkle-tree}
\includegraphics[scale=0.6]{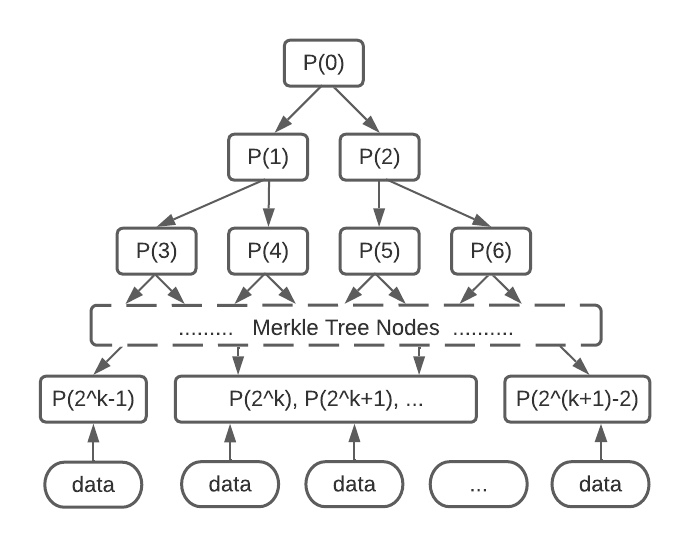}
\end{figure}

\subsubsection{State pinning.}
State Pinning\cite{robinson2019anonymous} is defined as including the state of one block-chain in another block-chain. In our scenario, we denote $h_k$ pins the state $s_k$ if $h_k$ is generated from a collision-resistant hash of $s_k$.
\\
\subsubsection{Side effect.}
We denote the changes of $tx_i^k$ on chain-state $C_k$ other than the partial state $s_k$ to be the side effect $e_k$ of of $tx_i^k$. Side-effects can be emitting events on native chains or native chain contract calls.
\\
\subsubsection{Polynomial Commitment schemes.}
Polynomial commitment schemes (PCS) provides the ability to commit to a polynomial over a finite field and prove its evaluation at points\cite{boneh2021halo}. A succinct PCS has commitment and evaluation proof size sublinear in the degree of the polynomial. An efficient PCS has sublinear proof verification. Any efficient and succinct PCS can be used to construct a SNARK with similar security and efficiency characteristics (in the random oracle model).
\\
\subsubsection{Zero knowledge proof of program execution.}
Since PCS provides a scheme to prove evaluation of polynomials at certain points, we can convert program evaluation into polynomials equations so that the problem of proving program has certain results can be converted into problems of polynomial equations. Further more, a zero knowledge proof of a program execution is a PCS proof that can be used by a prover to convince a verifier that a program is executed correctly with out leak any private information.\\
\newline
The main idea of our protocol is to introduce a new chain (we will call it {\bf aggregator chain} below) so that instead of executing transactions in a special order in a bunch of native block chains we do the following (see figure \ref{main-idea}):

\begin{figure}[!ht]
\caption{main idea of \dprotocol}
\label{main-idea}
\includegraphics[scale=0.4]{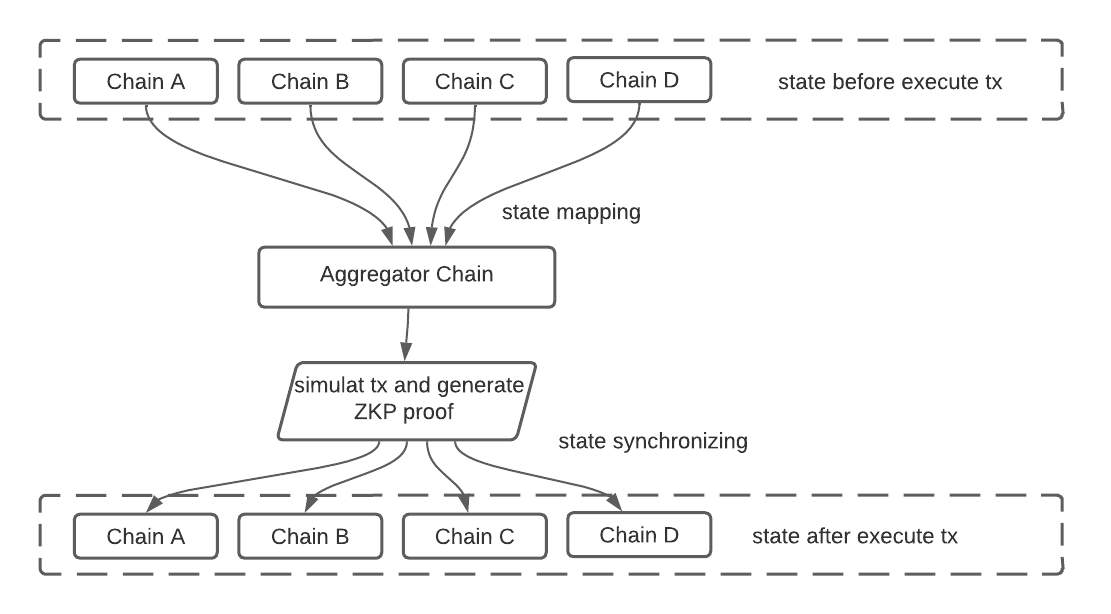}
\end{figure}
\begin{enumerate}[leftmargin=*]
\item We map local state $s_i$ of each block chain $C_i$ into our newly introduced aggregator chain to form a global state $s = \langle s_0, s_1, \cdots \rangle$.
\item We perform a simulating run of the bundled transactions over $s$ on aggregator chain to get a result state $s' = \langle s_0, s_1, \cdots \rangle$ and create zero knowledge proofs to convince native chain $C_i$ that after the bundled transaction has been executed, their local state has to be changed to $s_i'$.
\item We ask native chain $C_i$ to change their local state by providing them the result state $s_i'$ and the zero knowledge proof.
\end{enumerate}

Before elaborating more details of our solution, we would like to give a brief overview about a few existing cross-chain communication protocols and discuss their safety, permissionless, liveness, and linearization properties.

\subsection{Related Protocol Overview}
\subsubsection{Notary schemes:}

The main idea of notary schemes is to elect one or more trusted nodes as notary public and report transactions in different block-chain networks through notaries\cite{qin2018overview}. Therefore all information transferred between different block-chains are completely managed by notaries.

The centralized notary schemes is its efficiency in procession events and simplicity in implementation. However it suffers with single point of failure problem. Therefore, a multi-sig notary scheme may been proposed in order to reduce the trust on certain centralized node. However, this multi-sig notary are surely non-permissionless and needs extra protocols to ensure liveness.
\\
\subsubsection{SPV: Simplified payment verification}

    SPV is a special case of one direction state pining. In such protocals, block header of A and a merkle proof of a particular transaction was monitored and sent to target block-chain B. block-chain B accept the transaction trough calculating the partial merkle tree hash and compare it with the block header of block-chain A. The liveness properties of this approach depends on the liveness of monitors of A and the safety property relies on the blockheaders are reported to B honestly. 
\\
\subsubsection{Sidechains:}

    The concept of sidechains was first proposed in 2014, main goal of sidechain is to extend the scalability and functionality of the block-chain system. When using sidechain as an trusted centralized entity in a 2-way peg system, it can be used as a standalone transaction dealer. 
    Sidechain can enforce the security of transactions on itself by by implementing a protocol which can be validated by consensus. Since the sidechain needs to update state changes back to the underlying block-chains, the block-chains needs to trust or verify the transactions sent out by sidechains. The safety property again relies on whether observation of source chain can be honestly reported to the side chain and whether the verifier on target chain can reject all fraud transactions. In addition, since side-chain can suffer from deny-of-service attacks that leads to a bundled transaction can not be finalized, the safety property also relies on the liveness property.

    Liveness of communication based on sidechains relies on how robust the sidechain itself is and whether all the transactions happened on side chain will get reported to the target chain eventually. Some improving ideas are given in Plasma: Scalable Autonomous Smart Contracts for some improved ideas of sidechains.
\\
\subsubsection{Two way peg:}
    A 2-way peg (2WP) works like a two way SPV which allows the transfer of assets from one block-chain to another and vice-versa. The assets are technically not transferred, but temporarily locked on the source block-chain while the same amount of equivalent tokens are released in the target block-chain. The transferred asset can be withdraw when the equivalent amount of tokens on the target block-chain are locked again. The problem with this scheme is that the transfer is not finished until the target block-chain has release the equivalent amount asset. Therefore any 2WP system must do compromises and rely on assumptions about the honesty of the actors involved in the 2WP.
\\
\subsubsection{Cross Chain Gateway and relayers:}
    Cross Chain Gateway with relays is another extension of the idea of state pinning. While it enables block-chain interoperability applications including cross chain token transfer, the safety properties are got at the cost of storing every single block header of the source block-chain\cite{belchior2021survey}. In general the cost of storing such state is very expensive.
\\
\newline
In conclusion we have the following table:
\begin{table}[h]
\small
\centering
\begin{tabular}{ | c | c | c | c | }
\hline
Protocal & Liveness & Safety & Permissionless \\
\hline
Notary schemes & yes & yes & no\\
\hline
SPV & conditional & conditional & no \\
\hline
Peg & conditional & conditional & no \\
\hline
SideChain & yes & conditional & yes \\ 
\hline
Gateway& yes & yes but costy & no \\ 
\hline
\end{tabular}
\end{table}

\dprotocol derives the basic ideas from multi-way sidechain (multiple block-chains with a shared sidechain) and solves the trustworthy problem of the third parties by verifying the computation from sidechain via zksnark proofs. In \dprotocol, we denote our specific sidechain as aggregator chain and involved block-chains of bundled transactions as native chains.

Below We will present the brief picture of our solution by a case study of the standard transfer and explain how zero knowledge is applied so that the safety, liveness, permissionless properties hold. In Chapter 2 we describe each components in details and then we explain the trust base of our solution and prove that under the trust base our solution has full guarantee of safety, permissionless and liveness properties in Chapter 3 and 4. In chapter 5 we show that our solution has a linearization property. 

Moreover, since our permissionless properties relies on a BFT consensus algorithm which assumes that two thirds of the nodes of the aggregator chain are honest, we present a potential incentive strategy to keep enough decentralized nodes so that 2/3 majority attach is hard enough.

\section{A case study of cross chain transfer}
As a case study, we focus on a standard simplified asset transfer process from Alice on chain A to Bob on chain B through a liquid provider. 
\subsection{State abstraction}
Firstly we define the native-chain set in the bundled transfer transaction to be $\mathbb{C} = \left\{C_A, C_B\right\}$. As Alice has some asset on $C_A$ and Bob has some asset on $C_B$ and a liquid provide has asset both on $C_A$ and $C_B$, we can abstract our local state $s_A$ of $C_A$ and $s_B$ of $C_B$ separately as follows.

\begin{lstlisting}[escapeinside={(*}{*)}, basicstyle=\small, xleftmargin=0.6cm, backgroundcolor = \color{lightgray}, numbers=left]
(*$s_A$*) = state {
    alice: number
    lp: number
}
(*$s_B$*) = state {
    bob: number
    lp: number
}
\end{lstlisting}
Now, as we described in section \ref{prelimiary}, the underlying global state $s$ of transfer is a tuple of $C_A$ and $C_B$:
\begin{lstlisting}[escapeinside={(*}{*)}, basicstyle=\small, xleftmargin=0.6cm, backgroundcolor = \color{lightgray}, numbers=left]
s = state {
    sa: (*$s_A$*),
    sb: (*$s_B$*)
}
\end{lstlisting}
Secondly the bundled transfer transaction $tx$ can be defined as a function on global state $s$ as follows:
\begin{lstlisting}[escapeinside={(*}{*)}, xleftmargin=0.6cm, basicstyle=\small, backgroundcolor = \color{lightgray}, numbers=left]
function transfer(amount) {
  token1.transfer(Alice, amount, lp);
  
  ca.alice = ca.alice - amount;
  ca.lp = ca.lp + amount;
  
  cb.bob = cb.bob + amount;
  cb.lp -= amount;
  
  if(cb.lp > amount) {
    token1.transfer(lp, amount, Bob);
  }
}
\end{lstlisting}
where line 2 can be treated as the invoke transaction on $C_A$, line 4-12 are bundled transactions that will be simulated on aggregator layer and line 11 is the side effect on $C_B$.\\
\newline
Now, to execute the transaction safely we fit this transaction into \dprotocol and divide the life cycle of the transfer into four stages: consensus stage, operating state, proving stage and finalizing stage.\\
\begin{figure*}[!ht]
\caption{case study of a cross chain token transfer}
\label{case-study}
\includegraphics[scale=0.6]{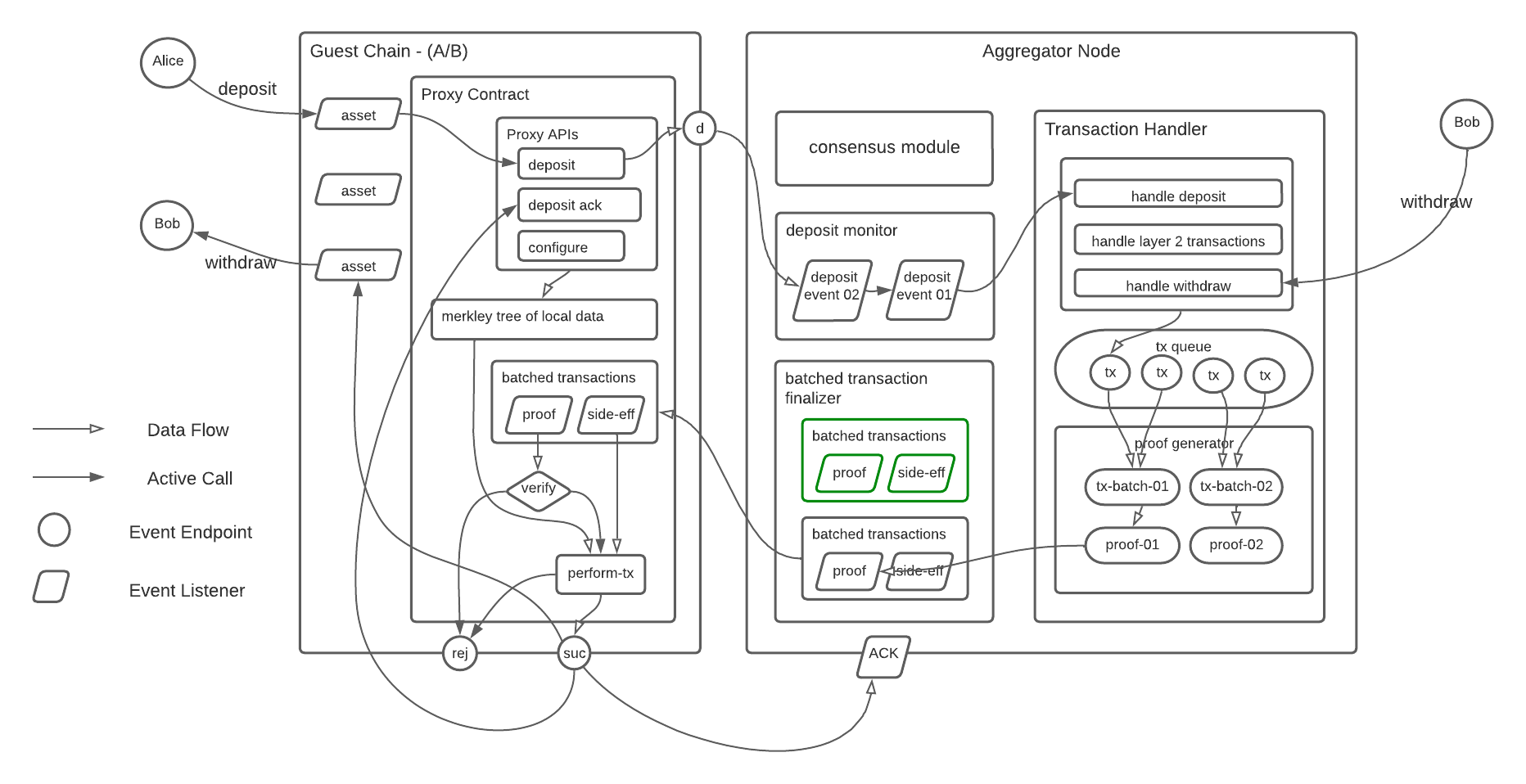}
\end{figure*}

\begin{enumerate}[leftmargin=*]
\item Consensus Stage:
User Alice transfer a certain amount of money into a proxy contract (line 2). After the contract handles the invoke transaction, it commits the amount into its table of unfinalized invoke transactions and emit a event. This emitted event will get monitored by Delphinus Aggregator Node and triggers the consensus stage.\\
\newline
In the consensus stage, a voted node will encoded the consensus algorithm so that it can prove (by ZKP) later to native chains that this invoke transaction was honestly monitored by at least two-thirds of the Delphinus Aggregator Nodes. \\
\newline
Also, Alice are allowed to withdraw unfinalized amount after a certain amount of time(based on blockheight) (see chapter 2 for more information about encoding consensus into ZKP circuits).\\
\newline
Invoke transaction are handled in this way because we need to make sure that if the transaction fails(at simulation stage and proving stage) then the whole transaction can be reverted by inverting the invoking transaction.\\
\item Simulating stage:
Once a transaction is ready to execute, it will be simulated by the voted aggregator chain node so that a new resulting global state $s$ can be calculated and by projecting $s$ on to different partial state, we know how to update the local state $s_A$ and $s_B$ on $C_A$ and $C_B$ accordingly.\\
\item Proving stage:
After Delphinus aggregator node finishes simulating the execution of our particular transfer transaction, it will create a zero knowledge proof to show that how to update all the partial state on $C_A$  and $C_B$.\\
\newline
Comparing to traditional side-chain solutions, who require audit nodes to validate transactions and may suffer from majority attacks, \dprotocol uses ZKP to enforce that the simulation is carries out on predefined functions (in this case, line 4-12 in transfer transaction). Thus even the majority of the audit nodes are hacked, they can only perform attacks that leads to deny-of-service of some of the transactions. Later in Chapter 4 we will prove that the DOS attack is also hard once we have a permissionless aggregator chain and providing ZKP(zero knowledge proof) for consensus algorithm as well.\\ 
\newline
Once the $tx$ was fully performed, aggregator chain will broadcasting this transaction together with its proof to native-chain $C_A$ and $C_B$.\\
\item Finalizing Stage:
After a zero knowledge proof our a simulation of our transfer is generated, the voted aggregator node will call a special finalize contract on native block-chain $C_A$ and $C_B$ which will verify the proof and then perform the side-effects of each sub transactions accordingly.\\
\newline
In the case of transfer, after the proof was verified, $C_A$ and $C_B$ will change their local state accordingly and on $C_B$, side effect (line 11) was performed so that Bob receives the transferred assert correctly(This amount must equal to the amount transferred by Alice other wise the zero knowledge proof will not be valid).
\end{enumerate}
\subsection{How zero knowledge proof technique is applied}
For correctness, the ZK verification on verify contract makes sure that all transactions are performed under predefined functions and the resulting state are correctly calculated (merkle root hash is pinned). The pinned amount makes sure that the custody of transferred tokens is reported to aggregator chain honestly and the transfer on $C_B$ is performed at end which makes sure that the amount about to withdraw is a consequence of valid transaction simulation.

\subsection{Cost of zero knowledge proof verification}
Regarding the cost problem of state pinning, since \dprotocol uses batched transactions to reduce the cost of signature check and snark proof check, the problems of the high cost of state pinning and signature check can be reduced to a acceptable level.

\section{Protocol Details}
The target of \dprotocol is to provide a universal layer to simulate and synchronise the global state between various block-chains which support smart contract. This universal layer is organized as a block-chain of Delphinus aggregator nodes and each node contains four abstract components: storage component, transaction handler and simulator, native chain relayer and transaction finalizer.  Besides aggregator nodes, we also need to install native proxy contracts on native block-chains so that they can invoke transactions and update local state accordingly.

\subsection{Delphinus aggregator chain node}
\begin{figure}[ht]
\caption{protocol components}
\label{protocol-components}
\includegraphics[scale=0.5]{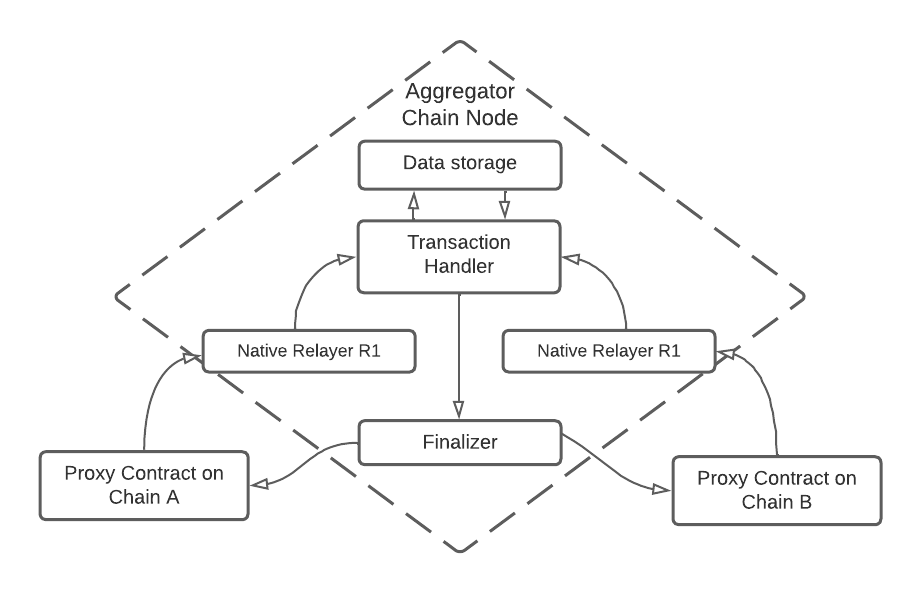}
\end{figure}

As an example, suppose that we have two block-chains $C_{A}$ and $C_{B}$, then, to perform cross chain transactions between $C_A$  and $C_B$, our single \dprotocol involves the following entities (figure \ref{protocol-components}). 

with each components performs the following functionalities:
\begin{itemize}
\item  a proxy smart contract $\mathcal{A}$ on $C_{A}$ (Trusted).
\item  a proxy smart contract $\mathcal{B}$ on $C_{B}$ (Trusted).
\item  a transaction handler that handles and simulates transactions.
\item  a storage component that tracks the global state.
\item  a relayer $\mathcal{R}_A$ to monitor invoke(revoke) transaction events for $\mathcal{C}_A$ and reports them to transaction handler.
\item  a relayer $\mathcal{R}_B$ to monitor invoke(revoke) transaction events for $\mathcal{C}_B$.
\item  a finalizer that communicates with proxy contracts to finalize transactions.
\item  a data storage component that tracks the global state.
\end{itemize}
Moreover, in a block-chain setting, we combine the finalizer, storage component, transaction handler and relayer (which are everything except proxy contract on native block-chain) together with an extra consensus component into a Delphinus aggregator chain node. These distributed Delphinus aggregator node forms a block chain that works together to synchronize and alter states between guest block chains and each components in node performs the following task:

\begin{enumerate}[leftmargin=*]
\item chain storage component:
    \begin{itemize}
    \item store the global state.
    \item store the merkle root hash of the whole state.
    \item store the valid voter list and the merkle root hash of all voters.
    \end{itemize}
\item transaction handler and simulator.
    \begin{itemize}
    \item order aggregator chain transactions under certain consensus.
    \item process transactions and update the whole state accordingly.
    \item emit transaction events (with execution order).
    \end{itemize}
\item native chain relayers:
    \begin{itemize}
    \item handle event emitted by invoke transaction of a bundled transaction.
    \item handle event emitted by revoke.
    \end{itemize}
\item transaction finalizer:
    \begin{itemize}
    \item calculating transaction execution proofs.
    \item calculating proofs of consensus.
    \item submit proofs to guest chain for validation and finalize.
    \end{itemize}
\item consensus component:
    \begin{itemize}
    \item performing consensus algorithm (voting block generator).
    \item calculate ZKP of consensus algorithm.
    \item generate and prepare block.
    \end{itemize}
\end{enumerate}

Last but not least, proxy contract on native-chain is also an important component of \dprotocol as it is used to maintain the partial state of a particular native-chain by performing proof verification, state updating and processing the sideffects of each transactions during the finalize stage.  

\subsection{Transaction lifecycle}
A bundled cross-chain transaction $tx$ in Delphinus aggregator node is a composed set of operations $tx_i^k: s_k \rightarrow s_k$. By defining how the local state $s_k$ of the application can be changed through $tx_i^k$, $tx$ can be treated as a function based on the global state $s$. In \dprotocol the global state of an application is compressed as a Merkle hash tree and pinned in all the guest chain proxy contracts.

\begin{figure}[!ht]
\includegraphics[scale=0.4]{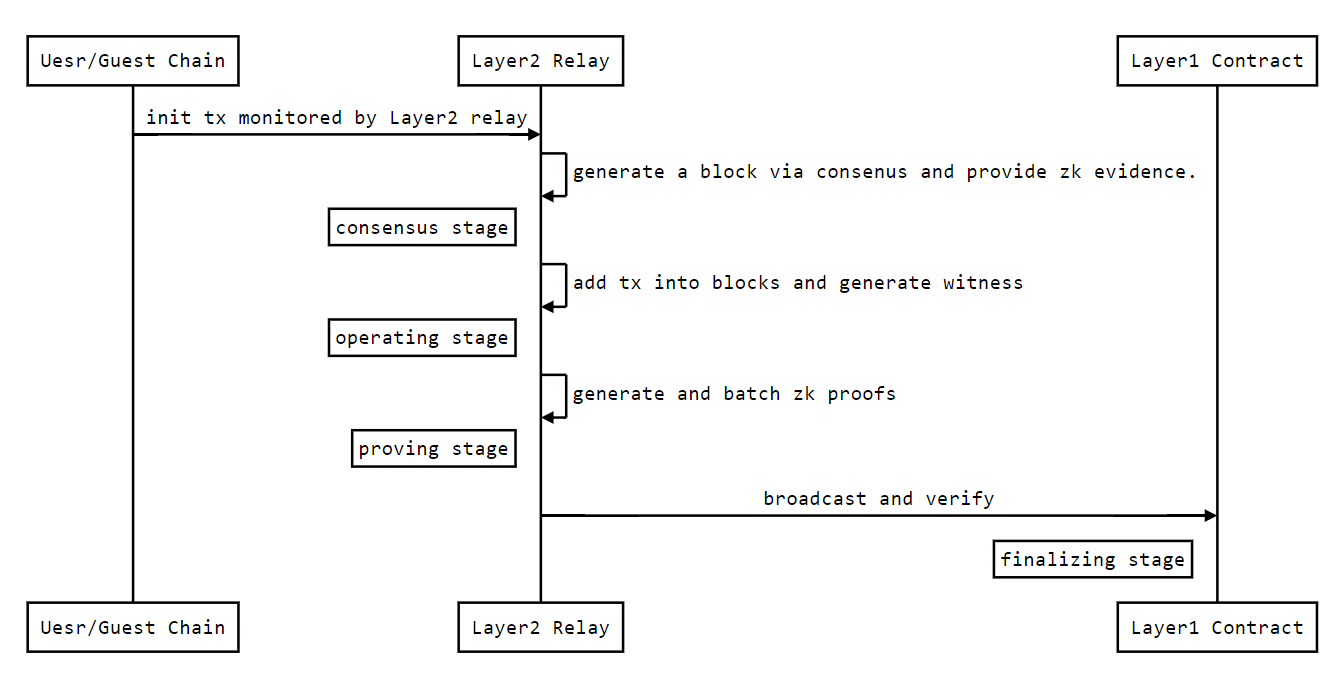}
\caption{transaction life cycle}
\label{consensus-sequence}
\end{figure}

For a transaction to be successfully executed in \dprotocol, it experience four stages: Consensus Stage, Operating Stage, Proving Stage and Finalizing Stage (see figure \ref{consensus-sequence}).
\begin{itemize}[leftmargin=*]
\item Consensus Stage: As a decentralized protocol, a transaction will not get handled until it is collected in a block. Each blocked is produced by an elected node under a pre-defined consensus algorithm (see 3.2). A node who wins the consensus game needs to provide a zk proof which will later be used as evidence to finalize all transactions within the block on guest proxy contract.

\item Operating Stage: At operating stage, an operation is simulated on Delphinus aggregator chain and changes the global state (S) thus changes the Merkle root in a sandbox. During the simulation, all the witness needed to create a zkp for it is prepared and stored. Since all transactions are handled sequentially regarding their order in a particular block, their witness are stored in the same order so that they can be batched in the proving stage. 

\item Proving Stage: After operating stage, the Delphinus aggregator chain relay will generate all the proofs for every transaction and compress proofs into a single zkp proof. This proof is then appended after the transaction arguments in the block with a updated merkle tree hash. 
\item Finalizing Stage: Once a proof of batched operation sequence is generated by Delphinus aggregator chain monitor, it will broadcast to various underlying block-chains. The block-chain needs verify the proof and then send acknowledge back to the Delphinus aggregator chain to finalize the transaction.
\end{itemize}

\subsubsection{Consensus Stage}
A consensus algorithm is a common process in block-chain to achieve agreement on value or state among distributed nodes. A consensus algorithm is critical in safety in systems that contain permissionless unreliable nodes. In \dprotocol we need consensus algorithm to solve the following problem:

\begin{enumerate}[leftmargin=*]
\item Who can decide the order of the bundled cross-chain transactions.
\item How to enroll and identify a valid node.
\item Who can produce the next block and finalize the transactions in it. 
\end{enumerate}

In \dprotocol we uses a modified PBFT\cite{castro1999practical} consensus algorithm between chain nodes to vote for a block generator. Recall that PBFT consensus consists of three phases: Pre-Prepare, Prepare, and Commit. At Pre-Prepare stage, the primary node is responsible for verifying the requests and generating corresponding pre-prepare messages. Then, the primary node will broadcast pre-prepare messages to all Replica nodes. After receiving the messages, Replica nodes will verify the legitimacy of those pre-prepare messages and then broadcast a corresponding prepare message.

\begin{figure}[!ht]
\begin{center}
\includegraphics[scale=0.5]{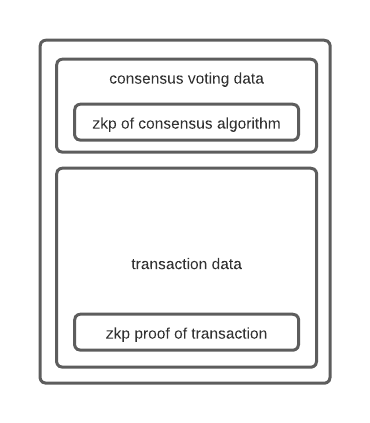}
\end{center}
\caption{data layout of a aggregator block}
\label{block-layout}
\end{figure}

In \dprotocol the final step is carried out on native chains thus the results are not broadcast back to replica nodes but broadcast to native chain nodes and verified there. Moreover, in addition to verifying that the results are correctly calculated we also need to convince guest chain that our generated block are produced under the consensus protocol as well. Usually since the native-chain do not attending the consensus algorithm in aggregator-chain, it has problems of checking whether the transaction results are reported from a valid picked node. Thus we need to encode the consensus algorithm and its result into a zero knowledge proof so that we can confirm guest proxy contract that the blocked which contains all transactions is generated legally. The consensus result together with its zkp proof is stored in a new generated block as described as follows (Fig \ref{block-layout}). 

\begin{figure}[!ht]
\includegraphics[scale=0.4]{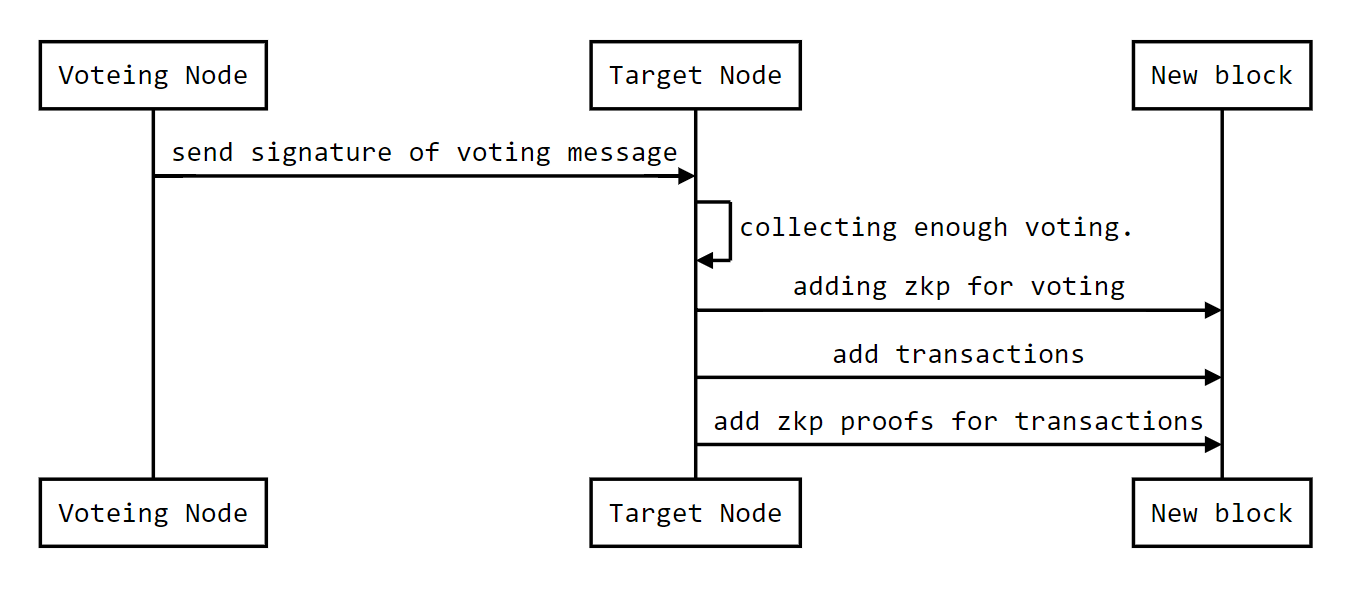}
\caption{voting process during consensus}
\label{vote-sequence}
\end{figure}

Thus after consider the above requirements, we used a modified PBFT approach which still consists of three phases but each phases is slightly modified as described below.\\
\newline
Firstly, at pre-prepare stage a node need to be registered as a valid block generating candidate and broadcast the ZKP of the validation to other nodes. During the process of voting (prepare stage), each aggregator chain node in consensus stage places a vote and send a signature of the vote to their voting target. Suppose that two thirds of the nodes in the system are honest then it follows that only one potential node will win the voting game and this particular node has the authority to produce a block.\\
\newline
Secondly, after a certain node gathers enough voting messages, instead of broadcasting the result, the winner node will calculate the ZKP(zero knowledge proof) of our consensus algorithm and put it into its generated block. This ZKP will later been used to convince native block-chain that a batched transaction proof is send from a legally voted node. Once the proof is attached in the newly generated block the winner node will start producing transactions(see figure \ref{vote-sequence}).

In the end, the winner node will broadcast the generated block to native block-chains and \dprotocol proxy contract on native block-chain will verify and finalize this block and update their local state accordingly.

In addition to produce zkp for consensus algorithm, \dprotocol also targets to build a permissionless chain that every one can join the aggregator chain to helping synchronizing bundled cross-chain transactions. This means we need to provide a protocol to register and remove a valid voter. To achieve that, \dprotocol keeps a merkle tree of all valid nodes by tracking their public keys as merkle tree data nodes and any one can register to be a valid node by registering itself as a new merkle tree leaf. Also, for a aggregator node to vote for the next block producer, it has to prove he is a valid merkle tree node thus a registered valid node first and then sign and send his voting ticket to the voted node. More precisely anyone can do the following:
\begin{enumerate}[leftmargin=*]
\item Register a node for voting:
    Node registration is a aggregator chain transaction that updates the merkle tree of all valid voters on aggregator chain and has side effect of updating the merkle tree root hash of nodes in guest proxy contract. It is every aggregator node's responsibility to keep the registration tree updated so that the root hash are consistent with the pinned hash in the proxy contract. \\

\item Valid a voter:
    When a node receive a voting from other node, it needs to validate that the sender is a valid voter. If the sender is an invalid node (eg, a malformed node without registration in) then proof of sufficent voter will fail at the stage of finalize thus waste the computing source of the node who is voted.\\

    Recall that all the valid voter is arranged in a merkle tree whose root hash is pinned in the guest proxy contract. The voter needs to provide a identity proof (in forms of zk proof) to show that his public key(or voting address) is one of the merkle tree leaf nodes.\\

    Once a node receive a voting from voter $V$, it needs to verify the identity proof of $V$ first before it push it to the list of his supporters for next block generation round.\\

\item Send a valid vote ticket:
    A vote ticket contains two parts, a signature and a validation proof. The signature is signed from the current block height together with merkely root hash of all valid voters. The validationg proof is a zero knowledge proof proves that the voter is one of the valid voters. 
\end{enumerate}
\subsubsection{Simulation Stage}
After a node is voted as a block generator, it can starting collecting transactions from various sources. In \dprotocol transactions are submitted to aggregator node via three sources(fig \ref{simulation-stage}).\\
\newline
\textit{1. Pure bundled transaction that does not contains a invoking transaction:} A pure bundled transaction is supposed to be send directly to \dprotocol nodes and all its internal sub transactions should either all succeed or fail. Also whether each sub transactions will succeed is decidable via simulation in aggregator node so that the aggregator node can simulate the result first and then finalize the state change on native block-chains.\\
\newline
\textit{2. Bundled transaction that contains a invoking transaction}: This kind of bundled transaction likes the pure bundled transaction except that its first sub transaction (and only the first sub transaction) can not be simulated by aggregator node. When handling this kind of transaction, the first transaction is executed on native-block chain first and the relayer of aggregator node monitors the result of the first transaction. If the first transaction is succeed, the relayer submits a following transaction to the aggregator node. If not, the relayer submits a revoke transaction to the aggregator which will trigger a revoke of the invoking transaction just performed on native-chain.\\
\newline
\textit{3. System transactions}: This kind of transactions are submitted by aggregator nodes and it contains voting transaction, voter registration transaction, voter quit transaction, etc.

\subsubsection{Proving Stage}
As we discussed before, we simulate bundled transactions of the global state on aggregator chain node and each bundled transaction is equivalent to a function $(f: state \rightarrow state)$ that alter and query the global state $s$. Since the global state is pinned on native-chain by the merkle hash root $h(s)$ of the state, a zero knowledge proof of bundled transactions $tx$ is a proof that ensures the merkle hash root of the result state $s' = f(s)$ is valid. 
\begin{figure}[!ht]
\begin{center}
\includegraphics[scale=0.8]{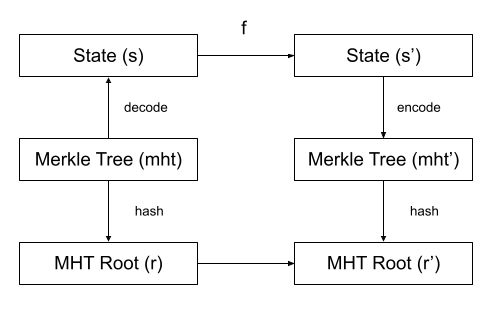}
\end{center}
\caption{state refine of transaction circuit}
\label{circuit-refine}
\end{figure}

For instance(see figure \ref{circuit-refine}), suppose that, before the transaction, state $s$ is encoded as $mht$ in a format of standard merkle tree which has root hash $r$ and after simulating the transaction the state changes to $s' = f(s)$. Also suppose that this new state was stored in the aggregator chain storage in a format of updated merkle tree $mht'$ whose new root hash is $r'$.
Then it follows that we need to create a zero knowledge proof to show that the following constraints holds.
\[ constraints = \begin{cases}
    r = root(mht(s)) \\
    r' = root(mht(s'))\\
    mht(s') = mht(f(s))
\end{cases} \]
where the first two constraints are constraints about the state before and after the transaction and the last constraint is about the simulation of function $f$.\\
\newline
Moreover, $f$ can be treated as a sequence of leaf update of the merkle tree of state $s$. For each leaf update, suppose that $P_k = [p_0, p_1, p_2 \cdots p_n]$ is the MTI of an updated leaf of data $D_k$. It follows that to calculate the root hash after the modification of data of $D_k$, we need to calculate new hash root at each level from $p_n$ to $p_0$ by the formula
$$
    H_k(p_k) = hash(H(p_{k+1}^0), H(p_{k+1}^1),\cdots) 
$$
where $a_i = p_{k+1}^{j}$ are nodes who have the same parent $p_k$ as $p_{k+1}$\cite{liskov2005updatable}. A full transaction $tx$ is a data transform function $f$ performed on multiple Merle leaf nodes. In Delphinus aggregator node $f$ is defined as a combination of the following:

\begin{enumerate}[leftmargin=*]
\item simple merkle tree leaf update.
\item predefined functions that has predefined zkp circuits.
\item crypto functions including signature functions and hash functions.
\end{enumerate}

\subsubsection{Finalizing Stage}
We know that a bundled transaction $tx$ is a state transform over global state $s$ and is composed of a sequence of sub transactions $tx^j_{k_j}$ on $C_{k_j}$. During the finalizing stage, \dprotocol performing mainly two things. Firstly it checks the ZKP of the consensus algorithm to make sure the block was generated by a valid block producer and then it checks the ZKP of the simulation of $tx$ so that the proxy contract can safely update the local state on $C_{k_j}$. Secondly the proxy contract performs the side effects of $tx^j_{k_j}$ on $C_{k_j}$ (e.g., if a user would like to withdraw some asset from aggregator chain to guest chain, then during the finalize procedure a transfer from guest chain proxy contract to the target account of the withdraw is executed).

\subsection{Guest Proxy Contract}
\begin{figure}[!ht]
\includegraphics[scale=0.15]{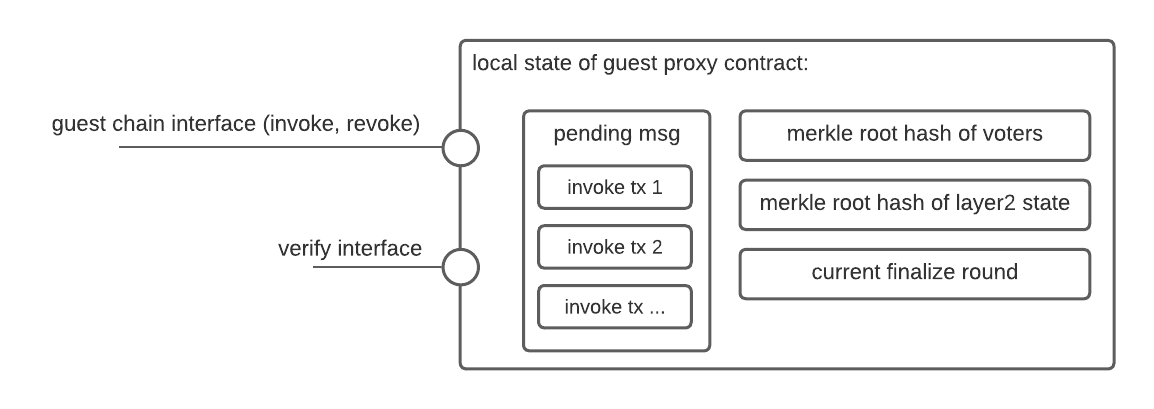}
\caption{local state of guest proxy contract}
\label{local-state}
\end{figure}

Guest Proxy Contract contains two parts: {\bf guest chain interface} and {\bf verification} interface (fig \ref{local-state}). Both of them manipulates the guest chain local state. Guest chain interface includes invoke and revoke which allows guest chain user to invoke or revoke a message. Verification interface is called by Delphinus cross chain node to finalize transactions.

\subsubsection{Guest chain interface}
When a user invoke a {\bf guest chain interface}, a transaction info with the current block height ($h$) is recorded in the pending queue of the guest local state. The user who performed invoke transaction can revoke his invocation any time when the transaction remains in the pending queue after a fixed amount ($\delta_h$) of block. This means Delphinus cross chain nodes have to notice and handle this event and finalize this event within $\delta_h$ blocks.

\begin{figure}[!ht]
\includegraphics[scale=0.125]{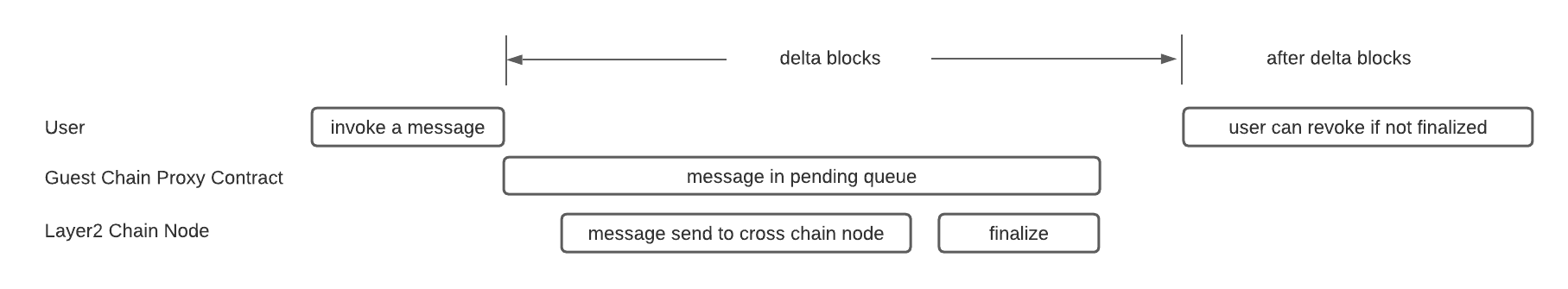}
\caption{time window for finalizing a block}
\label{case-study}
\end{figure}

\subsubsection{Verification Interface}
Verification interface verifies the ZK-SNARK proof generated by the aggregator chain node(see section 2.5).
\begin{code}
function verify(
  uint256 l2account,
  uint256[] memory tx_data,
  uint256[] memory verify_data, // zk proofs 
  uint256 vid,
  uint256 nonce,
  uint256 rid
)
\end{code}

Recall that only the winner of Delphinus voting consensus of the current finalize round (rid) have the authority to invoke verification interface at round $rid$. It follows that the verify function first check the voting proof to ensure that the sender is the winner of round $rid$ and then check every transaction encoded in tx\_data is executed honestly via verify the zk proof encoded in verify\_data. If both check pass, guest contract will perform all the side-effects of transactions on guest chain accordingly.\\

\subsubsection{Transaction side effects during finalize}
\begin{itemize}[leftmargin=*]
\item Invoke
    An invoke tx on aggregator chain must be related to a invocation call on guest contract. Since every time a guest chain invocation will cause a record of pending invocation to be added in to the pending list in the local state of guest contract.
    Once a invoke tx is verified, the side effect of that tx is removing the invocation record in the pending invocation list.
    Once a invocation was removed from the pending invocation list, its sender lose the capability to withdraw it.

\item Callback
    Callback is a special transaction that is invoked through Delphinus Cross Chain Node which tries to trigger a certain callback function from Delphinus Cross Chain to guest Chain.
    Once a callback tx was verified, its side effect is to calling a relataed guest contract api according to the tx info.
    Withdraw is a special case of callback which calles the trasfer api.

\item General Cross Chain Node Transaction:
    \begin{itemize}
    \item All general cross chain node transactions are handled on Cross Chain Nodes thus changes the cross chain state.
    \item As the cross chain state is pinned to guest contract there is no other side-effect during the process of finalization.
    \end{itemize}
\end{itemize}

\subsection{Guest Chain Relayer}
Guest chain relay is responsible for monitor guest chain events and notify aggregator chain tx handlers by forwarding them. Guest chain relay needs to sign the event with CrossChain Node's consensus id so that guest chain relay from another CrossChain Node could not forward invoking transaction to other's aggregator chain transaction handlers. If a voted CrossChain Node failed to finalize their transactions in guest chain prox contract, then it either failes to calculate a correct aggregated proof of transactions and voting result or it does not pass the checks in the side-effect of invoking transactions in guest contract. In later case, the failed node could be a malicious node who provides invoking transactions that does not exist on guest chain.  

\begin{figure}[!ht]
\caption{perform side effects of a bundled transaction}
\label{sideffects}
\includegraphics[scale=0.2]{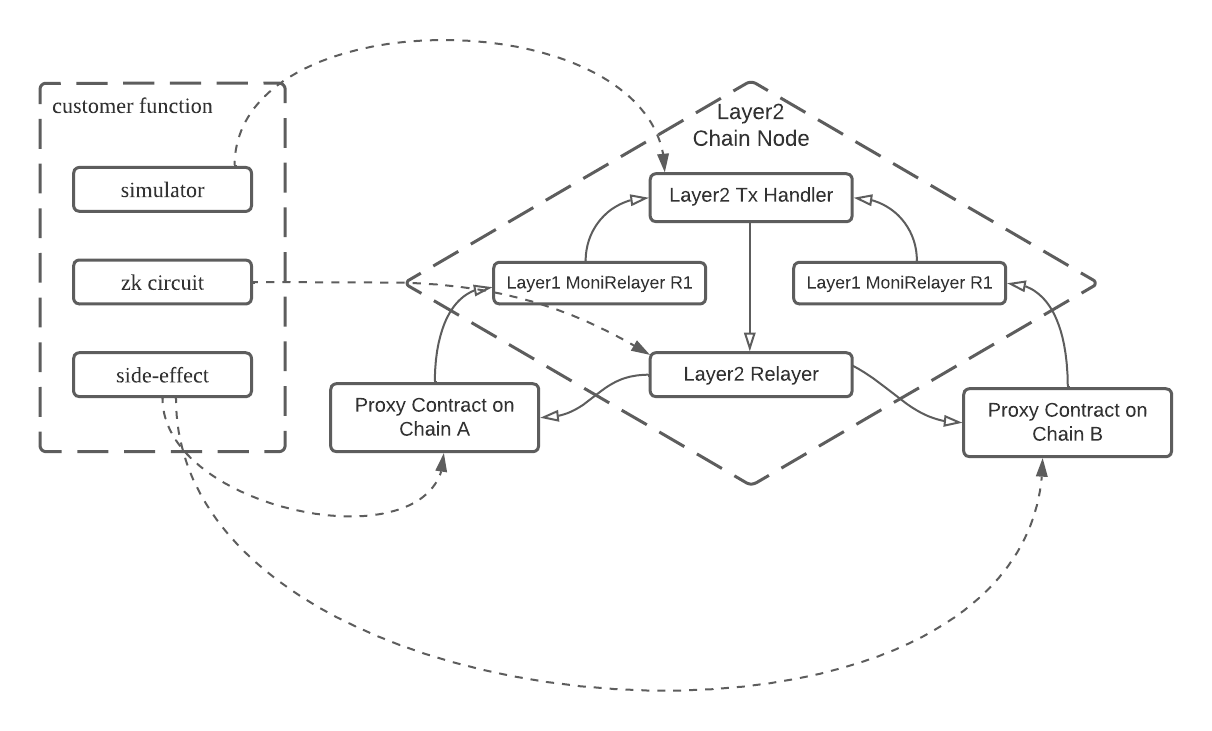}
\end{figure}

\subsection{Aggregate and broadcast multiple tx proofs}
Each time Delphinus aggregator chain relayer captures a full block generated by aggregator chain tx handler, it opens the block and generate a zk proof for each transaction and then aggregate the consensus proof together with all transaction proofs into one zk proof as the final proof of this block. After the final proof is added into the block, Delphinus aggregator chain relayer sends the final proof together with all the transactions contained in the block to delphinus proxy contracts for finalize.

zero knowledge proof aggregation is a zkp technique that can aggregate a sequence of proofs into one proof. The benifit of aggregating zk proofs is that it can reduce the calculation workload of verification. Recall that our aggregator chain relayer will generate a zk proof for each user transaction and each zk proof ensures that a transaction $tx_k$ changes state from $s_k$ to $s_{k+1}$ by proving that the root of merkely tree hash $H$ changes from $hash(s_k)$ to $hash(s_{k+1})$. An aggeregated proof batches all transaction proofs and generate a proof ensures that by execution a sequence of transaction $tx_0, tx_1, \cdots, tx_{n-1}$, the aggregator chain state changes from $s_0$ to $s_n$. 

Moreover, since the aggregator chain relayer needs to provide both the consensus proof and the aggregated transaction proof, we can further batch these two zk proofs into one.

\section {Permissionless and Liveness property}
\label{permissionless-liveness}

Delphinus cross chain is a decentralized platform that any one can join the network to help synchronizing cross chain bundled transactions. Thus to show that Delphinus cross chain is a permissionless system, we need to show that althought any running Delphinus cross chain node can play the role of synchronizing cross chain transactions, the data stored in Delphinus cross chain is consistent and all local data stored in distributed nodes converges.

\subsection{Proof of voting result}
Give a group of bundled transactions $\mathbb{B} = \left\{btx_i\right\}$ submitted simultaneously. A valid trace from a start point of global state $s$ is a sequence of state $s_0=s, s_1, s_2, \cdots s_n = s'$ such that $s_k = btx_k(s_{k-1})$ where $btx_k \in \mathbb{B}$.

It is obvious that different order of $btx_k$ from same $\mathbb{B}$ can leads to different valid traces which means all attending nodes in the Delphinus cross chain needs to vote for a unique block producer to generate a trace that is valid globally.

Each round before voting, every attending node's public key is registered in a merkley tree and the root hash is pinned in all native chains and nodes. We assume that each node will synchronize the root hash themselves before vote since if they vote with a wrong root hash then their vote ticket is ignored when producing the final zero knowledge proof of the voting result.

To vote a node, a voter needs to sign a voting message and send the signed voting message to the target it votes. Inside the message it contains the merkle root hash of all attendence. Once a node receives a sufficient number of voting tickets that carry the correct attendence hash it will produce a zeroknowledge proof $\mathbb{P}_v$ which shows that the total number of tickets reachs two-thirds of the total number of voters and all of them has the correct attendence hash. 

\subsection{Block generating}
As mentioned in 3.3, after the winner node calculated the zkp proof of all received signatures, it can start producing a block which represents a valid sequence of bundled transactions $btx_i$. This process is executed locally and the block producer can pick any order of $btx_i$ in the final sequence as long as it can generate a zkp proof of the simulation $btx_i$

\begin{figure}[!ht]
\begin{center}
\includegraphics[scale=0.4]{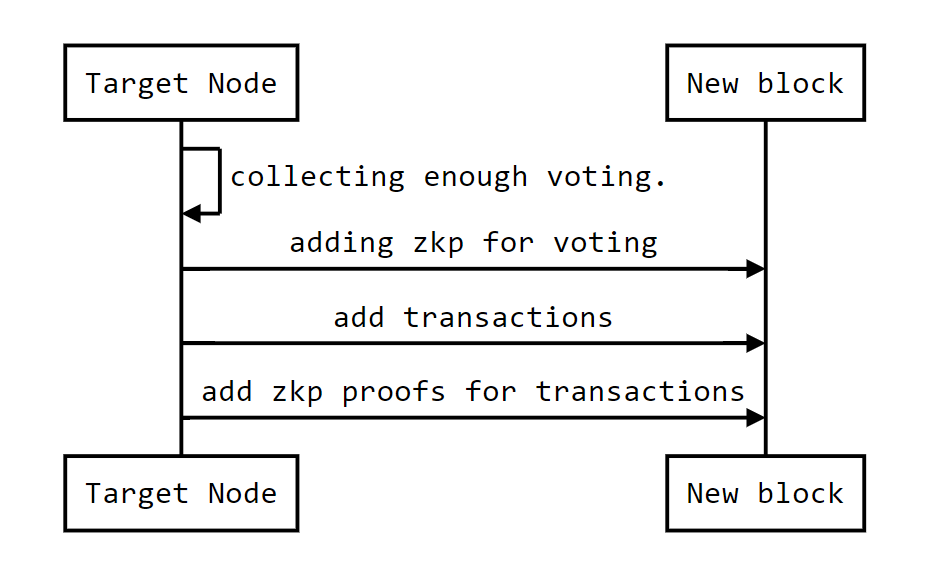}
\end{center}
\caption{produce a block after voting}
\label{produce-block}
\end{figure}

Once the block was generated, it needs to be send to the native block chain $C_i$ to finalize before it can be recorded in the Delphinus aggregator chain ledger. Since Delphinus nodes are permissionless, the liveness of the whole system should not relay on the liveness of a particular node. Thus we need to provide a protocal to make sure the winner node can not broadcast the block partially. For example if the winner only broadcast the block to a subset of the native block chains then we need a strategy to continue broadcasting the block even if the winner node quit after generating the block.

To solve the partial synchronize problem, we record the proof data on the native local chain $C_i$ so that if a block was synchronized partially, then all Delphinus node will notice the recorded proof and can continue broadcasting the proof until it is fully synchronized.

Moreover, if the winner does not broadcast its block to any of the local chains, then the voter need to trigger another vote to skip the current round of block producing and send the skip signal to all native block chains. Once all the native block received the skip signal, they will not accept any block with the current round number and the next round of block generation can start. If one of the native chain received the synchronizing call of the generated block, then all the nodes need to perform a revoke-skip vote so that they can revoke the skip signal on native block chains.

\subsection{Liveness property}
Based on protocol defined as above, we abstract the state transition diagram as follows:

\begin{figure}[!ht]
\label{state-transition}
\begin{center}
\includegraphics[scale=0.2]{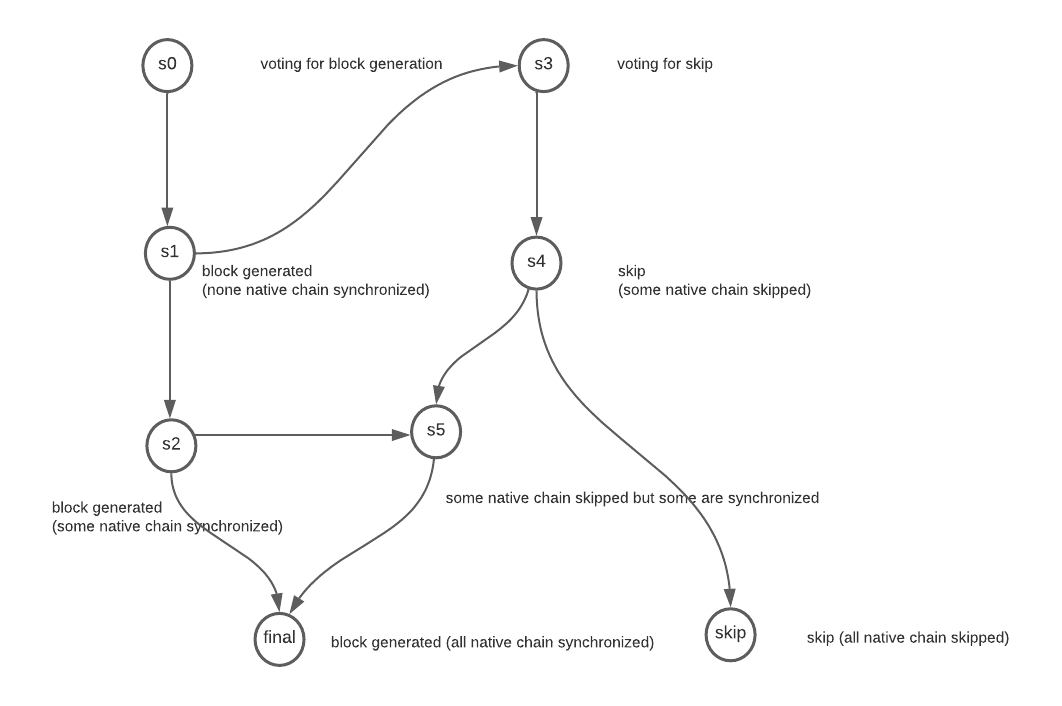}
\end{center}
\caption{state transition}
\end{figure}

Notice that in the protocol, there exists state $S_5$ in which both the block generation voting and skip voting are fired. If skip signal was fully applied to all native chains then the current round was skipped otherwise if any one of the native block chain receives the generated block, then our protocol ensures that all the other nodes will receive the same generated block as well. Thus the final state of all native-block chains converges to the same state.

\section{Safety property}
Recall that for a bundled transaction $tx = {tx_i^{k_i}}$. The safety property of a protocol to carry out $tx$ is that either all $tx_i^{k_i}$ are succeed or none of them succeed.

We split the proof of safety property of \dprotocol in to three steps. Firstly, we show that the safety property holds if a transaction $tx$ does not have an invoke transaction or any side-effects. Secondly we further prove that if all side-effects are safe then the safety property still holds. Finally we show that if the first transaction of a bundled transaction $tx$ is an invoke transaction, then we can rely on our liveness property to ensure that if the invoke transaction succeed then all following transactions will eventually be finalized and if the invoke transaction fail then it can be revoked using a revoke transaction so that the global state stays the same. 

\subsection{Bundled Transaction with no invoke transaction}
If a bundled transaction does not start with an invoke transaction, then it means the bundled transaction is invoked directly on the aggregator chain. Because all components $tx_i$ of $tx$ are simulated and performed on aggregator chain and a proof of the execution of $tx$ can not been generated if any $tx_i$ fails during the simulation, a transaction can only be executed a whole. Once the ZKP proof was generated for the simulation, it will be used to convince each native-chain to change their local state according to the result of the simulation.

If the proof is valid, and is submitted to native chains then it will pass the validation check and  then we expect that the local state of the native chain will changed accordingly with no exception. Since the liveness property promises that all native chain will native eventually receive the valid proof, it remains to show that the state updating protocol in our proxy contract is safe.

A smart contract is a piece of code that can not been changed once deployed, its behavior is fixed and predictable. Thus we can assume that our proxy contract always perform correctly according to its pre-defined functionality. Suppose that $s$ is the start global state, $tx$ is the transaction and $s'$ is the final state. When our proxy contract receives $MTH(s)$, $tx$, $MTH(s')$ and the state difference $\delta_{s}$ together with a zero knowledge proof $p(MTH(s), MTH(s'), tx)$, our proxy contract will do the following:

\begin{enumerate}[leftmargin=*]
\item Proxy contract checks that pined local global hash equals $MTH(s)$. This makes sure that the current local state $s_i$ is consistant with the global state $s$.

\item Proxy contract verifies the proof $p(s, s', tx)$. Once the verify succeed, it can guarantee that $s'$ is a valid final state after simulating $tx$ over $s$.

\item Compute the changes of $s_i$ and change the local state from $s_i$ to $(s_i)'$. Since updating local state is a monadic function of local state, it will never fail which means once the proof of $tx$ is verified, local state will be changed correctly for sure.

\item Proxy contract performs all the related side effects for each transactions $tx_k$ in bundled transaction $tx$. 
\end{enumerate}

We notice that step 4 is the only place that a finalizing call will fail even the proof are correct. To address this problem, we break down side-effects into two categories. The first category contains side effects like emit events, logging or pure history recording which are naturally safe since it will never fail by design. The second category includes unsafe function calls like assets transfer. This kind of unsafe functions can still be safe if we perform sanity checks in tx.

For example, if $tx_k$ in $tx$ will trigger a side-effect $e_k$ and $e_k$ is safe under condition $P$ where $P$ is a predicate of global state $s$, then we can insert a check after $tx_k$ as following:

\begin{code}
...
let s' = tx_k(s);
if !P_k(s'):
    raise Excepition(sainity check failure)
let s'' = tx_{k+1}(s);
...
\end{code}

Once $tx$ is modified so as above, a valid proof of a correct simulation of $tx$ will ensure not only the final state are valid but all the sainity checks are all valid and $e_k$ will be safe for sure.

\subsection{Bundled Transaction with an inovke transaction}
If a bundled transaction $tx$ starts with an invoke transaction $tx_0$ on $C_i$, then this invoke transaction on chain $C_i$ triggers the execution of the bundled transaction $tx$ by emit a invoking event. This event is reported to aggregator chain through an native-chain relayer. The aggregator chain bundled transaction simulator relies on the honesty of the reporter and if the relayer is hacked it can exploit the protocol by reporting malformed results of the inovke transaction. For example, in the transfer scenario, if a report reports the wrong amount of asserts that been transferred from Alice to Bob in Chain A then a successfully finalized $tx$ will trigger a wrong transfer from Bob to Alice on Chain B.

To address this problem and make sure the protocol is safe even the relayer is dishonest, we split the process of the changes of global state $s$ into two stages. At first stage $s$ is changed to $s'$ after invoke transaction and at second stage $s'$ is changed to $s''$ after the whole simulation of $tx$.

During the first stage $s'$ is pined in native chain $C_i$ once the invoke transaction is executed and the second stage is encoded by a zero knowledge proof generated from Delphinus aggregator chain  node. During finalize, proof of $tx$ now encodes the proof from $s'$ to $s''$. To ensure that the invoke transaction $tx_0$ is reported correctly, the verifier only needs to compare the MTH of local calculation of $s'$ and the hash provided by the aggregator chain Node. If the hash are the same, then it means $tx_0$ are reported honestly. More over since $s'$ are checked, $s''$ can be treated as a valid simulation from $s$ to $s''$. In conclusion the finalization steps are listed as follows:

\begin{enumerate}[leftmargin=*]
\item Proxy contract checks that pined local global hash equals $MTH(s'')$. This makes sure that the current local state $s_k^i$ is consistant with the global state $s'$ after processing invoke transaction. Thus the invoke transaction are reported honestly.

\item Proxy contract verifies the proof $p_{t}(s', s'', tx)$. Once the verify succeed, it can guarantee that $s''$ is a valid final state after simulating $tx$ over $s$.

\item Compute the changes of $(s^i)'$ and change the local state from $(s^i)'$ to $(s_i)''$. Since updating local state is a monadic function of local state, it will never fail which means once the proof of $tx$ is verified, local state will be changed correctly for sure.

\item Proxy contract performs all the related side effects for each transactions $tx_k$ in bundled transaction $tx$. 
\end{enumerate}

\section{Linearity property}
In concurrent programming, an operation (or set of operations) is linearizable if it consists of an ordered list of invocation and response events. In Delphinus cross chain protocol, the linearity property holds trivially. Because in each block generation round only one node wins the right to produce a block, the linearity hold iff and only if the final state simulated in the block can be represented as a sequence of bundled cross transactions $btx_i$. Since each block producer needs to create a zero knowledge proof to show that $s' = btx_k \circ btx_{k-1} \circ \cdots btx_0(s)$, a valid proof already shows the linearity property.

\bibliographystyle{plain}
\bibliography{main}

\end{document}